\documentclass[preprint, superscriptaddress, prb, footinbib]{revtex4}
\usepackage{graphicx}
\usepackage{amsmath}
\usepackage{color}

\begin{document}
\title{Elastic and vibrational properties of $\mathrm{\alpha}$ and $\mathrm{\beta}$-PbO.}
\author{P. Canepa}
\email{canepap@wfu.edu}
\affiliation{School of Physical Sciences, Ingram building, University of Kent, Canterbury CT2 7NH, United Kingdom}
\affiliation{Department of Physics, Wake Forest University, 1834 Wake Forest Road, Winston-Salem, NC 27109, USA}
\author{P. Ugliengo}
\affiliation{Dipartimento di Chimica, University of Torino, Via Pietro Giuria 7, Turin, 10125, Italy}
\author{M. Alfredsson}
\affiliation{School of Physical Sciences, Ingram building, University of Kent, Canterbury CT2 7NH, United Kingdom}
\keywords{Elastic constants, Phonons, Bulk moduli, EFG, DFT, DFT-D2, Layered materials}

\begin{abstract}
The structure, electronic and dynamic properties of the two layered $\alpha$ (litharge) and $\beta$ (massicot) phases of PbO have been studied by density functional methods. The role of London dispersion interactions as leading component of the  total interaction energy between layers has been addressed by using the Grimme's approach, in which new parameters for Pb and O atoms have been developed. Both gradient corrected and hybrid functionals have been adopted using Gaussian-type basis sets of polarized triple zeta quality for O atoms and small core pseudo-potential for the Pb atoms. Basis set superposition error (BSSE) has been accounted for by the Boys-Bernardi correction to compute the interlayer separation. Cross check with calculations adopting plane waves that are BSSE free have also been performed for both structures and vibrational frequencies. With the new set of proposed Grimme's type parameters structures and dynamical parameters for both PbO phases are in good agreement with experimental data.
\end{abstract}

\maketitle


\section{Introduction}
\label{sec:introduction}
Lead monoxide (PbO) is largely employed for several industrial and techological applications such as electronic devices,\cite{Pan2002,Sun2004,Murphy2006} in special ceramic-glasses,\cite{Takaishi2000,Fujino2004} for X-ray cathodes, for pigments,\cite{McCann1999,Pilania2009} in rubber vulcanization\cite{Heideman2005,Nanda2010} and in the automotive sector as an essential component for batteries.\cite{Cruz2002,Ahuja2011} PbO is largely found in two polymorhps: a tetragonal \textit{P4/nmm} phase ($\mathrm{\alpha}$-PbO or litharge) and an orthorhombic phase \textit{Pbcm} ($\mathrm{\beta}$-PbO or massicot).

\begin{figure}[!htbp] 
	\centering
		\includegraphics[scale=0.5,keepaspectratio]{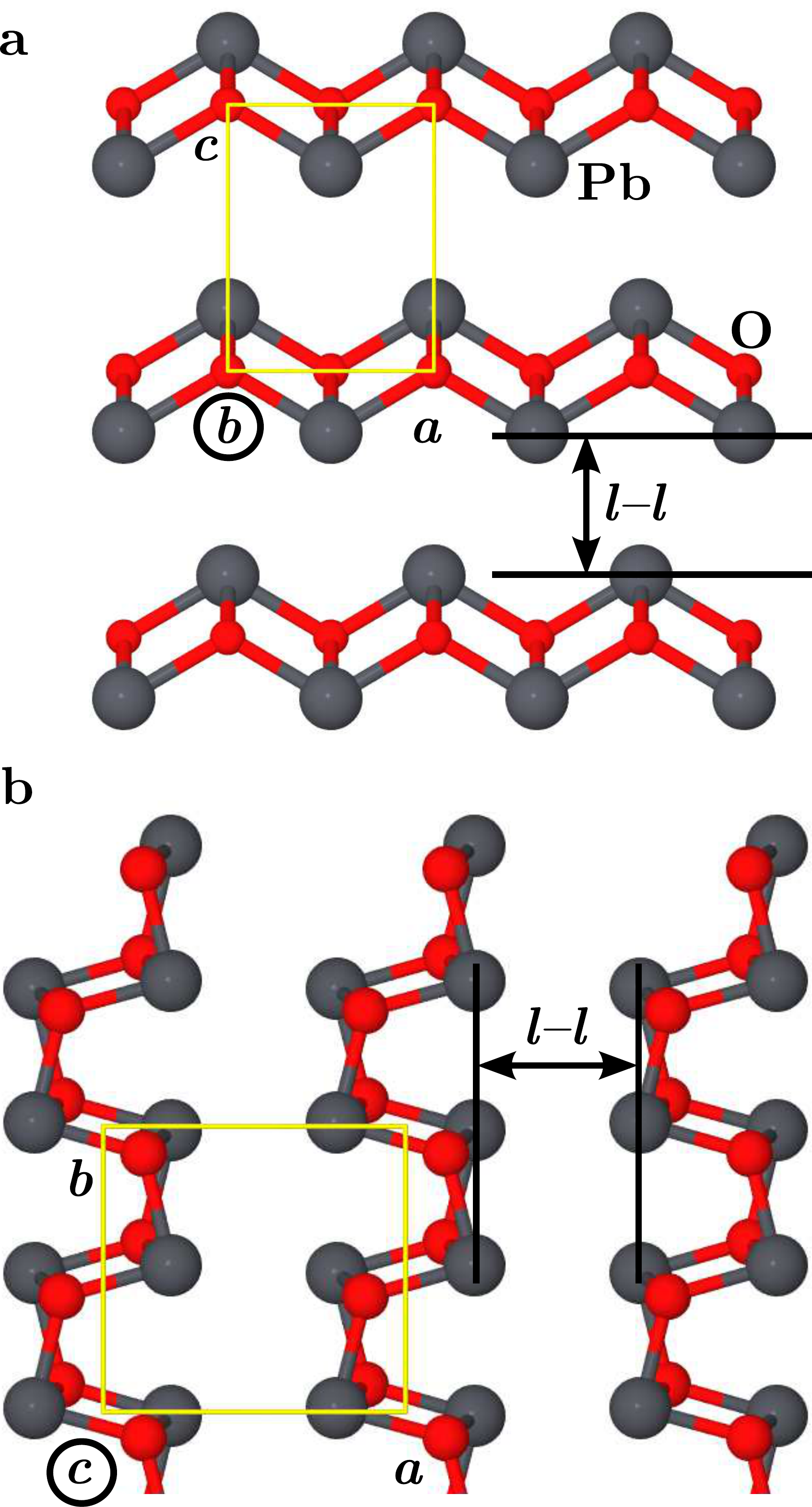}
	\caption{a) view along the [010] direction of $\mathrm{\alpha}$-PbO (\textit{P4/nmm}), and b)  along the [001] direction $\mathrm{\beta}$-PbO (\textit{Pbcm}). The inter-layer distance is highlighted as \textit{l-l}.}
	\label{fig:structures}
\end{figure}

In $\mathrm{\alpha}$-PbO the Pb$^{2+}$ ions are pyramidally coordinated by oxygen atoms (see Fig.~\ref{fig:structures}a) packed in a special layered-arrangement that resembles a distorted CsCl structure.
This distortion is principally caused by free lone-pairs on the Pb$\mathrm{^{2+}}$ ions. The key element in the lone-pairs localizations is the hybridization of the Pb(\textit{6s}) and Pb(\textit{6p}) orbitals with O(\textit{2p}) states as demonstrated by Watson \textit{et al.}.\cite{Watson1999a,Watson1999} Below 200 K the tetragonal $\mathrm{\alpha}$-PbO undergoes a phase transition to orthorhombic \textit{Cmma} as observed by Boher \textit{et al.};\cite{Boher1985} where the new \textit{a'} and \textit{b'} lattice parameters are redefined as $a^{\prime} \approx a+b$, $b^{\prime} \approx b-a$, where \textit{a} and \textit{b} are $\mathrm{\alpha}$-PbO lattice constants. The distorted $\mathrm{\alpha}$-phase can be regarded as intermediate PbO phase between $\mathrm{\alpha}$--and--$\mathrm{\beta}$ polymorphous. Very few studies are available on $\mathrm{\beta}$-PbO. This lead oxide polymorph is characterized by \textit{zig-zag} chains of PbO units repeated along the \textit{b} axis of the cell (see Fig.~\ref{fig:structures}b). As for litharge, Pb$^{2+}$ ions are found pyramidally-coordinated by oxygen atoms although, forming squares parallel to the \textit{bc} plane. The zig-zag chains are then stacked along the \textit{a} axis forming a layered structure almost unique in nature. While previous theoretical works established the structure of both PbO polymorphs\cite{Evarestov1991,Terpstra1995,Haussermann2001} and the nature of the Pb lone-pairs,\cite{Raulot2002} in this study we also address the role of dispersion interactions on the energetic, structure and dynamic properties of both phases. 

As depicted in Figure~\ref{fig:structures} in both lead oxide polymorphs rippled two-dimensional planes are stacked in a peculiar layered arrangement. The layered nature of $\mathrm{\alpha}$ and $\mathrm{\beta}$-PbO suggest that dispersive interactions may play a key role in stabilizing these structures. Differently from classic post Hartree-Fock methods such as MP2 and CCSD(T), most common DFT-GGA and hybrid functionals are unable to deal with purely dispersive London forces originating from fluctuating dipole-dipole interactions. Currently there are three main approaches to include dispersive forces in DFT:\cite{French2010} \emph{i}) the design of new functionals derived in a fully \emph{ab initio} fashion as suggested by Lundqvist and coworkers;\cite{Dion2004,Thonhauser2007,Lee2010,Vydrov2009} \emph{ii}) a highly parametrized functionals of the M0X family as propsed by Thrular \emph{et al.};\cite{Zhao2005,Zhao2008} \emph{iii}) an empirical correction to the standard DFT energy and gradient based on the empirical London formula as re-proposed by Grimme and later improved by Tkatchenko \emph{et al.}\cite{Grimme2004,Grimme2006,Grimme2010,Tkatchenko2009,Tkatchenko2009a} In this latter scheme, hereafter termed DFT-D2 (Grimme's correction),\cite{Grimme2004,Grimme2006} an atom-atom empirical potential (of the form $f(R)C_6/R^6$) accounts for the dispersion on the DFT energy as follows:
\begin{equation}
\label{eq:grimme}
E_{DFT-D2} =E_{DFT} + E_D  
\end{equation}

where $E_D$ is the additive dispersion term. 
\section{Computational Details}
\label{sec:method}
Simulations presented here were performed with CRYSTAL09,\cite{Dovesi2005,Dovesi2009} using hybrid and plain functionals. We used three GGA-functionals: PBE\cite{Perdew1996}, PBEsol\cite{Perdew2008}, BLYP\cite{Becke1988,Lee1988} and two hybrid-GGA functionals: B3LYP\cite{Becke1988,Becke1993,Lee1988} (20\% HF exchange) and PBE0\cite{Adamo1999} (25\% HF). Oxygen atoms were described with an all-electrons basis-set TPZ by Ahlrichs and co-workers (see Ref.~\citenum{Schafer1992}). Pb core-electrons were described by an effective small core potential\cite{Metz2000} along with a VDZ basis-set for the valence shells.\cite{Peterson2003} We used a Monkhorst-Pack\cite{Monkhorst1976} grid of 8x8x8 \emph{k}-points ensuring that the total energy was well converged. The self-consistent-field (SCF) procedure was converged within 10$\mathrm{^{-8}}$ Hartree. The Coulomb and exchange series were truncated using stringent overlap criteria, \textit{i.e.} 10$\mathrm{^{-9}}$, 10$\mathrm{^{-9}}$, 10$\mathrm{^{-9}}$, 10$\mathrm{^{-9}}$ and 10$\mathrm{^{-12}}$. Second-order elastic constants, $C_{i,j}$, were evaluated using the analytic total energy gradient and numerical second derivative with respect to the applied strain around the optimized equilibrium structure.\cite{Perger2009} According to the symmetry of the second-order elastic strain tensor, the appropriate number of strains were applied; hence the internal coordinates were relaxed for each strain displacement. Bulk moduli were determined \textit{via} the elastic constants (for tetragonal Voight averages: $\frac{1}{9}(2C_{11} + 2C_{12} + 4C_{13} + C_{33}$), and orthorhombic: $\frac{1}{9}((C_{11} + C_{22} + C_{33} ) + 2(C_{12} + C_{13} + C_{23}$)). For these calculations we reduced the SCF tolerance to 10$\mathrm{^{-9}}$ Hartree, whereas those on the gradient and the rms displacement were 6x10$\mathrm{^{-5}}$ and 1.2x10$\mathrm{^{-4}}$ Hartree bohr$\mathrm{^{-1}}$, respectively.\cite{Perger2009} The dynamical matrix, at $\mathrm{\Gamma}$ point,  was computed by finite differences: the atomic displacement was set to 3x10$\mathrm{^{-3}}$ \r{A}, reducing the SCF tolerance to 10$\mathrm{^{-11}}$ Hartree.\cite{Pascale2004} 

Aware of the spurious basis set superposition error (BSSE) introduced by the LCAO treatment employed by CRYSTAL09, we performed some PBE-D2 and -DC2 (see Sec.~\ref{subsec:dispersion}) calculations with a pseudopotential plane-wave (PP-PW) code \textit{PWscf} (a Quantum ESPRESSO package),\cite{Giannozzi2009} which is BSSE free. We used ultra-soft pseudopotentials for Pb (fully relativistic and with spin-orbit coupling correction) and O, whereas remaining valence electrons were described with a plane-wave basis set with a cutoff of 950 eV, while the total energy was sampled over a 8x8x8 k-point grid. Where not explicitly stated result were produced using CRYSTAL09. 

Structure manipulation and reppresentation were carried out with J-ICE.\cite{Canepa2011}

\section{Dispersive forces}
\label{subsec:dispersion}
The empirical dispersion contribution $E_D$, of Eq.~\ref{eq:grimme}, is defined as:
\begin{equation}
\label{eq:grimme2}
E_{D}=-s_6 \sum_{g} \sum_{ij} \frac{C_6^{ij}}{R_{ij,g}^6}f_{dmp}(R_{ij,g})
\end{equation}

where the summations extend over all atomic pairs $i, j$ and $g$ lattice vectors. $C^{ij}_6$ is the dispersion coefficient, and $s_6$ = 0.75 is a functional dependent scaling factor (see Ref.~\citenum{Grimme2006}). $R_{ij,g}$ is the inter-atomic distance between atoms $i$ in the reference cell and $j$ in the neighboring cells at distance $|g|$. All pairs farther than 25 \AA{} were disregarded in the summations due to their negligible contribution. Double counting for small inter-atomic distances are avoided using the following damping function $f_{dmp}(R_{ij,g}) = 1/(1+e^{-d (R_{ij,g}/R_{vdW}-1) }$) where $R_{vdW}$ are the atomic van der Walls' radii,  $d$ is the damping function steepness ($d = 20$).\cite{Grimme2004} In Grimme's work $R_{vdW}$ were set as the atomic van der Waals radii, which are 1.767 and 1.342 \AA{} for Pb and O respectively. The definition of the $C^{ij}_6$ coefficients of Eq.~\ref{eq:grimme4}\cite{Grimme2004,Grimme2006} follows the well-known geometrical mean:

\begin{subequations}
\begin{align}
C^{ij}_6 & = \sqrt{C^{i}_6C^{j}_6}; \label{eq:grimme4} \\
C^{i}_6 & = 0.05 NI^i_p \alpha^i. \label{eq:grimme5}
\end{align}
\end{subequations}

In Eq.~\ref{eq:grimme5}, $N$, is the number of the shell electrons, and has values of 2, 10, 18, 36, 54 and 72 for atoms from rows 1-6 of the periodic table. The original Grimme's $C^{i}_6$ parameters were derived from the atomic polarizabilites, $\alpha^i$, and ionization energies, $I^i_p$,\cite{Grimme2004,Grimme2006} leading to 63.16 for Pb and 0.7 for O $\mathrm{J nm^6 mol^{-1}}$, respectively. The $C^{i}_6$ for heavy elements (\textit{e.g.} Sn, Pb) were simply extrapolated from those of lighter elements of the same group,\cite{Grimme2004,Grimme2006} resulting in rather approximated values. Furthermore, the electronic nature (covalence or ionicity) of atomic species vary depending on the local chemical environment, imposing some limitations in the use of the atomic-like parameters. In order to give a better estimation of the C$^i_6$ values one should find a way to account for the different chemical environments of a given atomic species. Different methods to re-parametrize the $C^{i}_6$ for ionic systems have been proposed,\cite{Conesa2010,Allen2011} but, the $C^{i}_6$ values were derived \textit{ad hoc} for each system, reducing their transferability. In this study we derive the $C^{i}_{6}$ parameters for each ionic species in a \textit{ab initio} fashion following as close as possible the protocol suggested by Grimme by enforcing the role of the specific environment for each atomic species (details in the supplementary material). On elementary considerations, the net charges of the Pb and O ions should be +2 and -2, respectively. Mulliken analysis, albeit quite dependent on the adopted basis set, gives a much reduced values of +1 and -1, respectively. Nevertheless, L\"owdin charges calculated using a PP-PW approach agree with the Mulliken's picture. Using a pragmatic approach we compute the ionization potential and polarizability for the bare Pb$^+$ and Pb$^{2+}$ ions using a Stuttgart ECP with a QVZ basis set. The same methodology cannot be adopted for computing these quantities for O$^{-}$ and O$^{2-}$, since both species are unstable with respect to the free atom. In the latter case we have adopted a method proposed for MgO by Tosoni and Sauer\cite{Tosoni2010} to set up a proper environment for oxygen in order to get both O$^-$ and O$^{2-}$ as stable species. As described in the supplementary information we average the values of the polarizabilities over the two Pb$^+$/Pb$^{2+}$ and O$^{-}$/O$^{2-}$ states to account for the semi-ionic nature of the PbO oxides. Since the ionization potentials are intrinsically discontinuous variables we choose to adopt the values for the Pb$^{2+}$ and O$^{2-}$ to be used in the definition of the dispersion coefficient C$^i _6$.  Within this assumption the final $C^i_6$, with Eq.~\ref{eq:grimme5}, for Pb and O, become 36.93 and 0.54 Jnm $\mathrm{mol^{-1}}$, respectively (see supplementary Information). The new $C^i_6$ are smaller than the atomic-like ones reported by Grimme. The geometrical mean of the single $C^i_6$ parameters (see Eq.~\ref{eq:grimme4}) results in a mean $C_6$ (PbO) equal to 4.48 Jnm $\mathrm{mol^{-1}}$, which is found smaller than that proposed by Grimme, \emph{i.e.} 6.64 Jnm $\mathrm{mol^{-1}}$, avoiding the occurrence of spurious over-binding effects. 
In summary two flavors of DFT-D2 were employed: \textit{i}) using the Grimme's parameters (PBE-D2) and \textit{ii}) using the recalculated $C^{i}_6$ (PBE-DC2) according to the scheme presented here. Results are shown in Table~\ref{tab:geometry}. 

\section{Results and discussion}
\label{sec:results}
Results are outlined through three thematic sections: in Table~\ref{elegeom} we address the description of the geometrical properties of the PbO phases, while discussing the effect of dispersion on these materials. This section terminates with an insightful investigation of the lone-pairs nature using the electric filed gradient.  Sec.~\ref{elastic} discuss the elastic properties, while Sec.~\ref{phonon} vibrational frequencies.

\subsection{Geometry}
\label{elegeom}

Both $\mathrm{\alpha}$ and $\mathrm{\beta}$-PbO polymorphs crystallize in a layered arrangement.\cite{Terpstra1995,Watson1999,Watson1999a,Haussermann2001,Raulot2002,Waghmare2005} The PbO layered stacking and the interlayer distances (\textit{l-l}, see Fig.~\ref{fig:structures}), are controlled by lone-pairs on the Pb-sites. The resulting deformed electron cloud (of $\mathrm{Pb^{2+}}$) produces an electric dipole that along with dispersion forces collectively drive the layers to stack. Structures, of Fig.~\ref{tab:geometry}, were obtained after full structural relaxation (at 0 K) from experimental X-ray data of the PbO phases.\cite{Wyckoff1963,Hill1985}

\begin{table}[!h]
 \caption{lattice parameters \textit{a}, \textit{b} and \textit{c} (in \r{A}), \textit{c/a} ratio, Volume (in \r{A}$\mathrm{^{3}}$) and interlayer \textit{l-l} distance (see Fig.~\ref{fig:structures}, in \r{A}) of $\mathrm{\alpha}$, $\mathrm{\beta}$-PbO. Data in bracket are BSSE corrected.}
 
\label{tab:geometry}
\centering 

\begin{tabular*}{1\textwidth}{@{\extracolsep{\fill}} l c c c c c  } 
\hline
\hline
      $ \mathrm{\alpha}$-PbO \textit{P4/nmm} & \textit{a} &  \textit{c} & V & \textit{c/a} & \textit{l-l}  \\
  \hline
      Exp.\textsuperscript{\emph{a}} & 3.975 &  5.023 & 39.7 & 1.263  & 2.6   \\
       PBE\textsuperscript{\emph{b}}& 4.097 &  5.096 (5.600) & 41.9  & 1.221(1.367)   & 3.0   \\
       PBE-PW\textsuperscript{\emph{c}} &  3.985 & 5.579 & 45.8 & 1.400  &   3.2  \\ 
       PBE-D2\textsuperscript{\emph{b}} &  4.049 & 4.584 (4.766) & 37.6 &  1.132 (1.180) & 2.4 \\
       PBE-D2-PW\textsuperscript{\emph{c}}& 3.985 & 4.762 & 37.8 & 1.195  &  2.4\\ 
       PBE-DC2\textsuperscript{\emph{b}} & 4.066 & 4.648 (4.848) & 38.4  &   1.143 (1.192)&  2.5 \\
       PBE-DC2-PW\textsuperscript{\emph{c}} & 3.980&  4.835 &  38.9 &   1.215& 2.5    \\
       PBEsol\textsuperscript{\emph{b}} & 4.035  &4.690  & 38.2& 1.162& 2.3   \\
       PBE0\textsuperscript{\emph{b}}  & 4.021  & 4.996 & 40.4&  1.242  & 2.6 \\ 
       BLYP\textsuperscript{\emph{b}}&  4.144 & 5.746 & 49.3&  1.386 &  3.3  \\
       B3LYP\textsuperscript{\emph{b}} &  4.073  & 5.691 &47.2  & 1.397  & 3.3  \\
       LDA-PZ\textsuperscript{\emph{d}} &3.956& 4.874 & 38.1 & 1.232  &  -  \\
       LDA-PZ\textsuperscript{\emph{e}}  &3.953 & 4.988 &- & -& -\\
       LDA-PZ\textsuperscript{\emph{f}}&3.910&  4.930 & - & -& -\\

\hline
	   $\mathrm{\beta}$-PbO \textit{Pbcm} & \textit{a}  &  \textit{b} &\textit{c} & V  & \textit{l-l}   \\

\hline
       Exp.\textsuperscript{\emph{g}}& 5.893&   5.490  & 4.753 & 153.8 & 3.2   \\ 
       PBE\textsuperscript{\emph{b}} &  6.213 &  5.587 &4.823 & 167.4  &     3.5\\
       PBE-PW\textsuperscript{\emph{c}} &  6.293 &   5.805 & 4.800 & 147.9  &      3.6  \\ 
       PBE-D2\textsuperscript{\emph{b}}  & 5.499  &   5.324       & 4.732  & 145.3&  2.8 \\
       PBE-D2-PW\textsuperscript{\emph{c}}& 5.541 &  5.388 & 4.846 &  144.7 &   2.8\\
       PBE-DC2\textsuperscript{\emph{b}}   &   5.688 & 5.413   & 4.836  & 148.9 &  2.9 \\
       PBE-DC2-PW\textsuperscript{\emph{c}}& 5.636 & 5.445  & 4.836 & 148.4   & 2.9   \\
       PBEsol\textsuperscript{\emph{b}}  &  5.962  & 5.302   &  4.762 &  150.5&   3.2  \\ 
       PBE0\textsuperscript{\emph{b}}  &6.174   & 5.586  & 4.704 &  162.2   &  3.4   \\ 
       BLYP\textsuperscript{\emph{b}}  &  6.554 &   5.868 & 4.879 & 187.7  &    3.8  \\
       B3LYP\textsuperscript{\emph{b}} & 6.471   &  5.798 &  4.791 &179.8 &  3.7\\
       LDA-PZ\textsuperscript{\emph{f}}   & 5.770  &   5.380 & 4.680 & -& - \\

\hline
\hline
\end{tabular*}
  
\textsuperscript{\emph{a}}Ref.\cite{Wyckoff1963} X-Ray; \textsuperscript{\emph{b}}This work, LCAO; \textsuperscript{\emph{c}}This work, PP-PW; \textsuperscript{\emph{d}}Ref.\cite{Watson1999,Watson1999a} PP-PW; \textsuperscript{\emph{e}}Ref.\cite{Raulot2002} PP-PW; \textsuperscript{\emph{f}}Ref.\cite{Terpstra1995} augmented spherical-waves; \textsuperscript{\emph{g}}Ref.\cite{Hill1985} X-Ray.
\end{table}

Table~\ref{tab:geometry} shows how the LCAO method reproduces with good accuracy both the experimental\cite{Wyckoff1963,Hill1985} and previous LDA-DFT data.\cite{Terpstra1995,Watson1999,Watson1999a,Haussermann2001,Raulot2002,Waghmare2005} With PBE we found the $\mathrm{\alpha}$ phase more stable than $\mathrm{\beta}$ one consistently with experimental evidences ($\mathrm{\Delta E}$ = 3.22 kJ $\mathrm{mol^{-1}}$ per  formula unit). On the other hand the distorted $\mathrm{\alpha}$ phase, observed at low temperature, is negligibly more stable (at the PBE level) than the $\mathrm{\alpha}$ one for about 0.040 kJ $\mathrm{mol^{-1}}$ per formula unit.

Among the GGA functionals PBE is by far more accurate than BLYP; the latter largely overestimates the \textit{l-l} distance together with the lattice parameters \textit{a} and \textit{c}. Eventually B3LYP and PBE0, further increase the magnitude of the lattice parameters (see Table~\ref{tab:geometry}). $\mathrm{\alpha}$-PbO turns into  $\mathrm{\beta}$-PbO at 4.15 GPa with PBE (2.80 GPa with LDA), while experimental data ranges from 3.0 up to 3.6 GPa.\cite{Giefers2007} From this preliminary evaluation PBE appears to be the best choice and therefore it will be used to discuss the dispersion effects throughout the paper. Non-surprisingly functionals based on Becke exchange (BLYP and B3LYP) show an over-repulsive behavior compared to Perdew's functionals (PBE, PBE0) in agreement to Ref.~\citenum{Wu2001}.

\subsubsection{The effect of the dispersion}
\label{sec:bsse}
\begin{figure}[!htbp]
  \begin{center}

      \includegraphics[scale=0.7,keepaspectratio]{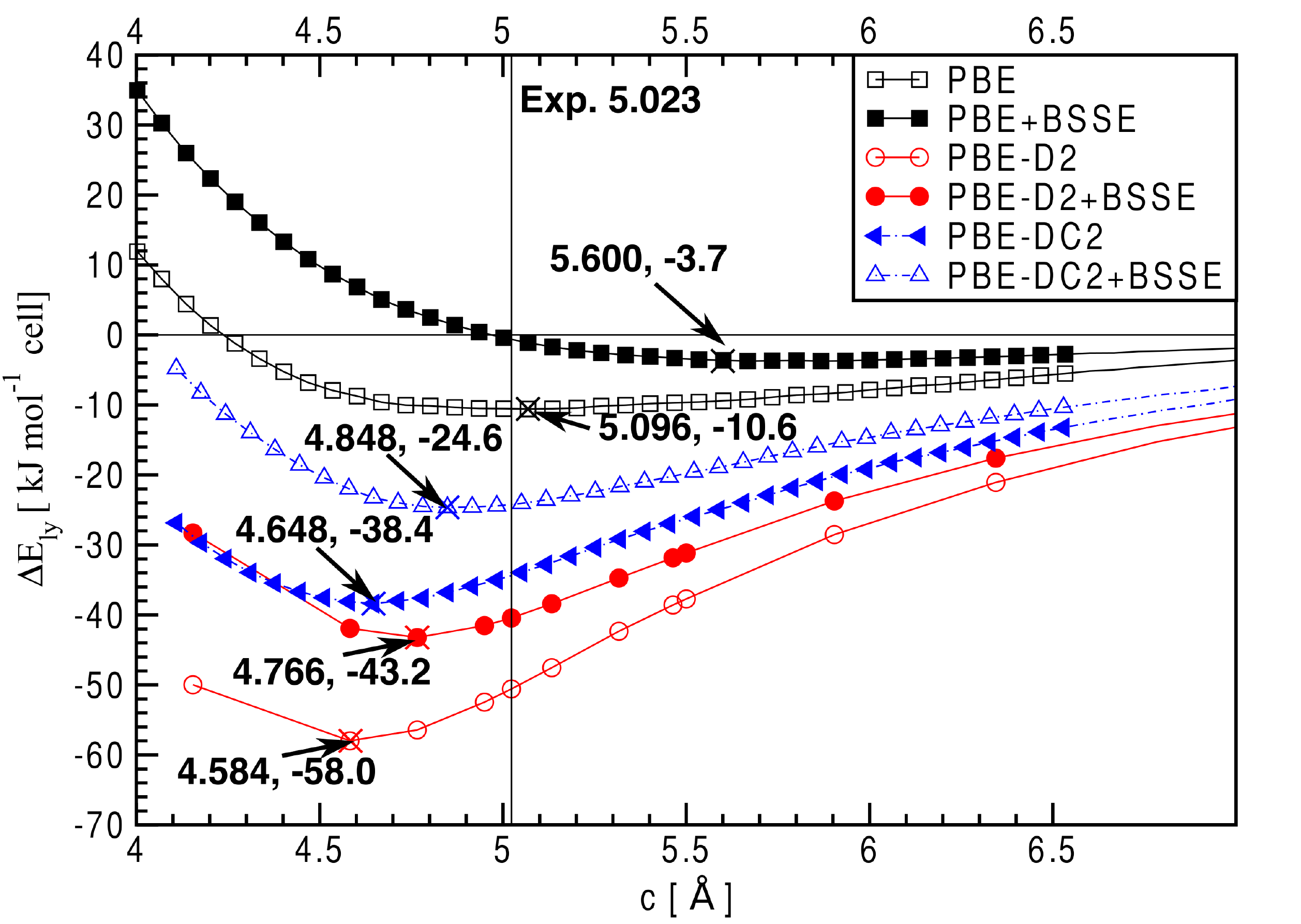}

    \caption{$\Delta E_{ly}$, and $\Delta E_{ly}^C$ \textit{vs.} \textit{c} for $\mathrm{\alpha}$-PbO using the PBE, PBE-D2 and PBE-DC2. $\Delta E_{ly}$ in kJ $\mathrm{mol^{-1}}$ and \textit{c} in \AA{}. The current graph is shortened at 7 \AA{}, but SCF calculations were run up to 40 \AA{} (\textit{c}) where $\Delta E_{ly}$ is null. }
    \label{fig:Fig_be}
  \end{center}
\end{figure}

While dispersion is accounted for both phases, the detailed discussion only concerns $\mathrm{\alpha}$-PbO. Insights on the layered structure of PbO is given by $\Delta E_{ly}$, which determines the extent of the interaction between two PbO sub-layers: 

\begin{equation}
\label{eq:deltaly}
\Delta E_{ly}=E_c-E_{\infty}
\end{equation}

where $E_c$ is the energy of two PbO layers, with atoms in the same geometrical relationships as in the bulk structure, at a given inter-layer distance (see \textit{l-l} in Fig.~\ref{fig:structures}) and $E_{\infty}$ is the energy of two PbO layers well separated and non-interacting. $\Delta E_{ly}$ is the energy cost of extracting a PbO sheet from the bulk (see Fig.~\ref{fig:structures}). Practically, this is done by running several SCF points at increasing lattice parameters (\textit{i.e. c} for $\mathrm{\alpha}$-PbO), eventually affecting the \textit{l-l} distance between 2 PbO layers. The real effect of the dispersion contribution, introduced by DFT-D2 or DFT-DC2, should only affect the $\Delta E_{ly}$ magnitude. Fig.~\ref{fig:Fig_be} shows the behavior of $\Delta E_{ly}$ at increasing lattice parameters \textit{c} in $\mathrm{\alpha}$-PbO.

In Figure~\ref{fig:Fig_be}, $\Delta E_{ly}$ changes dramatically when the D correction is introduced (see both PBE-D2 and PBE-DC2 cases). The PBE curve presents a very shallow minimum of -10.6  kJ $\mathrm{mol^{-1}}$. The correction introduced on $E_{ly}$ by PBE-D2, -58.0 kJ $\mathrm{mol^{-1}}$, is of 47.4 kJ $\mathrm{mol^{-1}}$ with respect to plain PBE. The PBE-DC2 data lies between the PBE and the PBE-D2 curves (-38.4 kJ $\mathrm{mol^{-1}}$, see Fig.~\ref{fig:Fig_be}). The empirical dispersion term, E$_D$ of Eq.~\ref{eq:grimme} and~\ref{eq:grimme2}, to the total energy promotes the inter-layer interaction forcing a decrease in the interlayer spacing. In $\alpha$-PbO the reduction of the \emph{c} lattice constant is the clear evidence of an increase in the layer-layer interaction.  The final magnitude relies totally on the size of the $C^i_6$ parameters that enters Eq.~\ref{eq:grimme2}. The large binding contribution of PBE-D2 and the PBE-DC2, is affected by the BSSE. The BSSE has a two-fold effect: \textit{i}) it reduces the relative $\Delta E_{ly}$ (\textit{i.e.} shifting the binding-curve to more negative $\Delta E_{ly}$ values over-binding PbO layers), \textit{ii}) artificially reduces the magnitude of \textit{c}. The BSSE was calculated using the counterpoise correction, and was practically done by introducing ghost functions on the two PbO layers while they were progressively separated. Knowing the magnitude of the BSSE one can re-calculate the corrected dispersion curve, whose minimum results shifted from the non-corrected one. The PBE BSSE corrected \textit{c} value is 5.600 \AA{}, which is largely overestimated with respect to the experimental value (see ~\ref{tab:geometry}, data in parentheses). Calculations with the PBE functional using \textit{PWscf} (BSSE free) confirmed our findings with the LCAO method.  

Summarizing, the PBE functional largely overestimates the lattice parameters involved in the stacking process. All LCAO data of Table~\ref{tab:geometry} when corrected for BSSE are shifted to larger values. PBE-D2 over-binds the $\mathrm{\alpha}$-PbO structure (\textit{c} = 4.584 \AA{}, see Table~\ref{tab:geometry}), whereas when corrected for the BSSE, the \textit{c} value (4.766 \AA{}) moves towards the experimental value (5.023 \AA) but still underestimated. This is confirmed by the PP-PW calculations. On the other hand, the re-parametrization of Grimme's coefficients introduces a visible improvement in the geometry description of $\mathrm{\alpha}$-PbO. The \textit{c} value (after correcting for the BSSE) is found in better agreement with the experimental value (4.848 \AA{}). This also agrees with PP-PW calculations (\textit{c} =  4.835 \AA{}). With PBE-DC2 $\mathrm{\alpha}$-PbO is more stable than the $\mathrm{\beta}$ phase by 3.6 kJ $\mathrm{mol^{-1}}$ per formula unit and also comparable with the PBE value 3.2 kJ $\mathrm{mol^{-1}}$ per formula unit. This shows that dispersion is very similar for the two PbO phases. Although we did not calculate the $\Delta E_{ly}$ for the $\mathrm{\beta}$ polymorph a very similar behavior of Fig.~\ref{fig:Fig_be} is expected. In the next sections only data obtained with PBE and PBE-DC2 will be considered, disregarding the PBE-D2.  

Although the reproduction of band-gap is not appropriate with DFT, the introduction of the dispersion reduces the band-gap simply through the decrease of the the \emph{c} parameter, which for $\mathrm{\alpha}$-PbO is 2.9 eV with PBE and it becomes 2.8 eV PBE-DC2 and 2.2 eV with PBE-D2. As observed by Allen \textit{et al.}\cite{Allen2011} the band-gap decrease is concomitant with the reduction of the \textit{c} parameter. This is the only effect noticeable by looking at the band structure and density of state plots (not reported here).

\subsubsection{Electric field gradient and quadrupole coupling constants}

\begin{figure}[!htbp]
\begin{center}

      \includegraphics[scale=0.7,keepaspectratio]{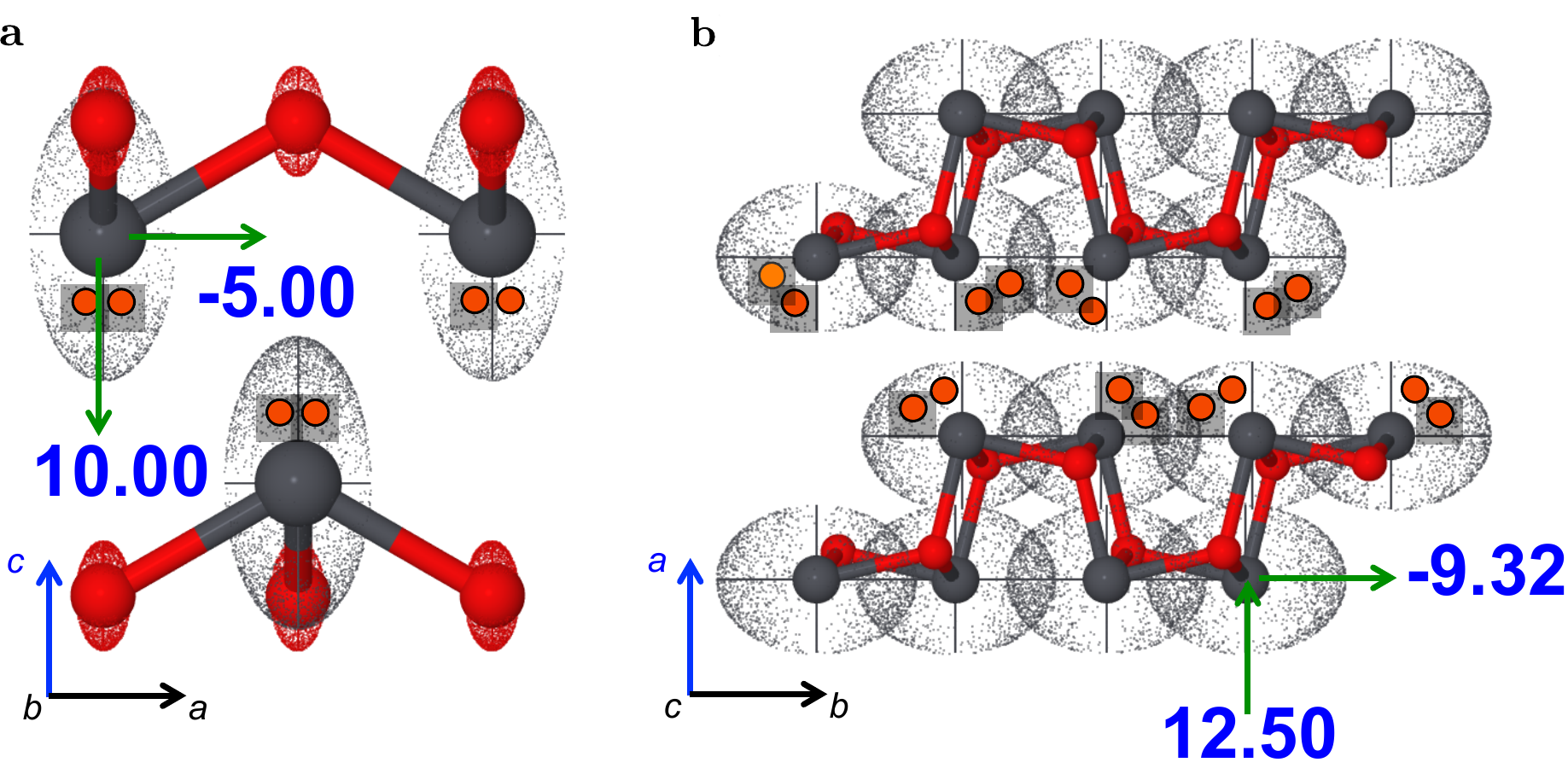}
	\caption{EFG principal components (arrows) superimposed on the two PbO structures, a) $\mathrm{\alpha}$, b) $\mathrm{\beta}$, respectively. $V_{11} $, $V_{22}$ and $V_{33}$ refers to data in Table~\ref{tab:efgqcc} and are in 10$\mathrm{^{-1}}$ $\mathrm{e}$ $\mathrm{a.u.^{-3}}$. Orange circles represent the lone-pairs within the interlayer zone. }
	\label{fig:Fig_efg}
	\end{center}
\end{figure}

The electric field gradient tensor (EFG) is beneficial to understand how the PbO lone-pairs arrange within the interlayer space (see distance \textit{l-l} in Fig.~\ref{fig:structures}).\cite{Kaufmann1979} EFG components: $V_{11}$, $V_{22}$ and $V_{33}$ are ordered according to their magnitudes  $V_{11} \leq V_{22} < V_{33}$. Relevant is the EFG asymmetry, $\eta$, calculated as $\mathrm{\eta = \frac{\vert V_{22} \vert - \vert V_{11} \vert}{\vert V_{33} \vert}}$, whereas the quadrupole coupling constant (QCC) is computed from $V_{33}$:

\begin{equation}
\label{eq:qcc}
QCC = \frac{e^2 q_m Q}{h} = \frac{eQ V_{33}}{h} 
\end{equation}

with $e$ the electron charge and $Q$ the atomic quadrupole moment. EFG tensor components, $\eta$, and QCC values for $\mathrm{^{17}O}$ and $\mathrm{^{204}Pb}$ are shown in Table~\ref{tab:efgqcc} and Fig. ~\ref{fig:Fig_efg}. 

\normalsize
\begin{table}[!htf]
\caption{PBE EFG components, $\eta$ and QCC for the following nuclei: $\mathrm{^{17}O}$ and $\mathrm{^{204}Pb}$ of $\mathrm{\alpha}$ and $\mathrm{\beta}$-PbO. Z for nucleus. $V_{xx}$ are expressed in 10$\mathrm{^{-1}}$ $\mathrm{e}$ $\mathrm{a.u.^{-3}}$ (1 $\mathrm{e}$ $\mathrm{a.u.^{-3}}$ $=$ $9.717x10^{21}$ $\mathrm{V m^{-2}}$), QCC in MHz.} 
\centering 
\begin{tabular*}{1\textwidth}{@{\extracolsep{\fill}} l c c c c c } 
\hline\hline 
Z &  $V_{11} $ & $V_{22}$ & $V_{33}$ & QCC & $\eta$\\
\hline
 \multicolumn{6}{c}{$\mathrm{\alpha}$-PbO \textit{P4/nmm}} \\
 Pb\textsuperscript{\emph{a}} & -5.00 & -5.00 & 10.00 & 104.17 &  0.00 \\
  O\textsuperscript{\emph{b}}  & -1.88 & -1.88 & 3.76   & 1.13  & 0.00 \\

\hline
\multicolumn{6}{c}{$\mathrm{\beta}$-PbO \textit{Pbcm} } \\
  Pb\textsuperscript{\emph{a}}   & -3.16  & -9.32 & 12.50 & 130.21 & 0.16 \\
  O\textsuperscript{\emph{b}}  &  -0.49 &  1.14 & 2.14 & 0.64  & 0.06\\

\hline\hline
\end{tabular*}
\label{tab:efgqcc} 
 \textsuperscript{\emph{a}}Ref.\cite{Friedemann2004} experimental Q($\mathrm{^{204}Pb}$) = 4.42x10$\mathrm{^{-29}}$ m$\mathrm{^2}$; 
\textsuperscript{\emph{b}}Ref.\cite{Pyykko1989} experimental Q($\mathrm{^{17}O}$) = -2.56x10$\mathrm{^{-30}}$ m$\mathrm{^2}$.
\end{table}

$V_{33}$ in $\mathrm{\alpha}$-PbO confirms that the lone-pair is oriented along the \textit{c} axis as previously observed with electron localization functions (ELF) plots.\cite{Raulot2002} The lone-pair of Pb atoms in the $\beta$-phase is found at an angle of 125.9, 38.2 and 101.5 degrees with respect to lattice constants \emph{a}, \emph{b} and \emph{c}. Differently from the $\alpha$-phase in $\beta$-PbO the Pb lone-pairs are not entirely oriented in one direction and similar evidences were discussed by Rault \textit{et al.}\cite{Raulot2002} Friedemann \textit{et al.}\cite{Friedemann2004} affirmed that the QQC for the $\beta$-polymorph (158.96 MHz) is larger than the corresponding QCC (96.82 MHz) value in the $\mathrm{\alpha}$-phase, which agrees with our results. Our \textit{ab initio} data is also confirmed by LAPW simulations ($\mathrm{\beta}$-QQC 151.55 $\mathrm{\alpha}$-QCC 93.55 MHz, respectively).\cite{Friedemann2004}  Le Bellac \textit{et al.}\cite{Le1992} also observed that the phase transition $\mathrm{\alpha \longrightarrow \beta}$ is accompanied by an evident change in orientation of the lone-pair. PBE-DC2 EFG components are equal to those calculated with plain PBE showing that dispersion interactions are too weak to alter the component of the electron density at nuclei.
\subsection{Elastic properties}
\label{elastic}

Table~\ref{tab:elastic} shows the elastic constants and bulk moduli of $\mathrm{\alpha}$ and $\mathrm{\beta}$-PbO calculated with the PBE and PBE-DC2 functionals.

\normalsize
\begin{table}[ht]
\caption{PBE, PBE-DC2 elastic constants $C_{ij}$ and bulk moduli B (calculated \textit{via} elastic constants), in GPa, for $\mathrm{\alpha}$ and $\mathrm{\beta}$-Pb. DC2 refers to PBE-DC2.} 
\label{tab:elastic}
\begin{center}
\begin{tabular*}{1\textwidth}{@{\extracolsep{\fill}}l c c c c c c c c c c }
	\hline
	\hline
& $C_{11}$  & $C_{22}$   & $C_{33}$ & $C_{44}$ & $C_{55}$ & $C_{66}$ & $C_{12}$ & $C_{13}$  &   $C_{23}$  & B\\
\hline
	 \multicolumn{11}{c}{$\mathrm{\alpha}$-PbO \textit{P4/nmm}} \\
PBE\textsuperscript{\emph{a}} &	 64.7    &  64.7          & 16.3  &  10.9  & 10.9 & 54.4 &  64.6 &   14.7 & 14.7  & 36.9  \\	
PBE-DC2\textsuperscript{\emph{a}}&        53.2     &  53.2        &   30.2 &  18.7  &  18.7 &     44.2       &   51.3  &  20.0 &   20.0    & 35.5\\
LDA\textsuperscript{\emph{b}} & -        &      -          &  -      &    -    &  -    &   -  &      - &  -      &    -    &    24.0    \\
Exp.\textsuperscript{\emph{c}}& - & -& -& -&- & -& -& -& -&23.1 \\
\hline 
	  \multicolumn{11}{c}{$\mathrm{\beta}$-PbO \textit{Pbcm}} \\
PBE\textsuperscript{\emph{a}} &  42.1 & 45.5  & 96.6 & 39.8 & 7.8 & 1.8 & 6.9  &  11.9 & 27.0 & 30.6 \\
PBE-DC2\textsuperscript{\emph{a}} &  48.4  & 47.3 & 102.2 & 30.6 &  22.5&  2.9 &   7.2 &  12.8 &  25.4 &  32.1\\
LDA\textsuperscript{\emph{d}} & - & -& -& -&- & -& -& -& -& 31.1\\
	\hline
	\hline

\end{tabular*}

\end{center}
\textsuperscript{\emph{a}}This work, LCAO; \textsuperscript{\emph{b}}Ref.\cite{Watson1999,Watson1999a} PP-PW; \textsuperscript{\emph{c}}Ref.\cite{Giefers2007} Exp.; \textsuperscript{\emph{d}}Ref.\cite{Haussermann2001} PP-PW. 

\end{table} 

Bulk moduli of Table~\ref{tab:elastic} are very similar to previous LDA simulations and experimental value in the case of $\alpha$-PbO,\cite{Watson1999,Watson1999a,Haussermann2001,Giefers2007} confirming the soft nature of these materials. Previous LDA simulations\cite{Watson1999,Watson1999a} behave significantly better than our PBE results. Bulk moduli for the $\mathrm{\beta}$-phase are in closer agreement with the LDA data. Although the calculated elastic constants are consistent with the geometries of the PbO-phases, the experimental values are currently not available. For $\mathrm{\alpha}$-PbO $C_{11}$, $C_{22}$ (64.7 GPa) are larger than $C_{33}$ (16.3 GPa) suggesting that the distortion along the $\mathrm{[001]}$ direction is easier (see Table~\ref{tab:elastic}), which agrees with the layered-structure. The inverse trend is obtained for $\mathrm{\alpha}$-PbO shear stresses. The effect exerted by the lone-pair on $\mathrm{\beta}$-PbO is smaller than in the $\mathrm{\alpha}$-phase; in fact $C_{11}$ (42.1 GPa) acting orthogonally to the $\mathrm{[100]}$ direction (\textit{i.e.} the \textit{a} direction) is similar to $C_{22}$ (45.5 GPa), which acts along the zig-zag chains. The strain along the $C_{33}$ (96.6 GPa) remains the hardest one according to the structural arrangement. Mixed strains ($C_{12}$, $C_{13}$, and $C_{23}$) and pure shear stresses ($C_{44}$, $C_{55}$ and $C_{66}$) are consistent with the geometry definition of both lead oxide-phases. We observed that PbO macroscopic densities increase when the $\mathrm{\alpha}$ phase is irreversibly transformed into its $\mathrm{\beta}$ one. Bulk moduli and elastic constants calculated with PBE-DC2 are similar to PBE. PBE-DC2 improves the bulk modulus of $\mathrm{\alpha}$ PbO toward the experimental value.

\subsection{Phonon frequencies at the $\Gamma$ point}
\label{phonon}
$\Gamma$ point phonon frequencies were computed by using PBE and PBE-DC2. The  relevant infrared (IR), Raman frequency window for both monoxides, is relatively narrow: 100-500 cm$\mathrm{^{-1}}$.\cite{Adams1976} This is likely to cause overlapping between near bands as confirmed by Adams \textit{et al.}.\cite{Adams1976} $\alpha$-PbO, with space group \textit{P4/nmm} ($D_{4h}^{7}$) gives rise to 9 vibrational modes (see Eq.~\ref{eq:gammaalpha}). The PbO-$\mathrm{\beta}$ phase \textit{Pbcm} ($D_{2h}^{11}$) shows 21 vibrational modes (see Eq.~\ref{eq:gammabeta}).

\begin{subequations}
\begin{eqnarray}
\Gamma^{\alpha} & =  &A_{1g} + A_{2g} + B_{1g} + B_{2g} + 2E_{g} +\nonumber \\
          &  + & A_{1u} +  A_{2u} + B_{2u} + 2E_{u} \label{eq:gammaalpha}
\\
\Gamma^{\beta} & = &3B_{2u} + 4B_{1g} + 4A_{g} + 2B_{2g} \nonumber \\
         &+ & 3B_{3u} + 2A_{u} + B_{1u} + 2B_{3g} \label{eq:gammabeta}
\end{eqnarray}
\end{subequations}

Vibrational frequencies calculated within the LCAO approximation are intrinsically affected by the BSSE error. To understand the magnitude of the BSSE on the final result we have compared the IR/Raman frequencies computed with Gaussian basis-set calculations with those obtained with a PP-PW approach (\emph{i.e.} \emph{PWscf}). Results reported in the supplementary material show a good agreement between the two dataset implying that BSSE does not dramatically affect the vibrational frequencies. 
\subsubsection{$\mathrm{\alpha}$-PbO} 
\begin{figure}[!ht]
	\centering 
	\includegraphics[scale=0.62,keepaspectratio]{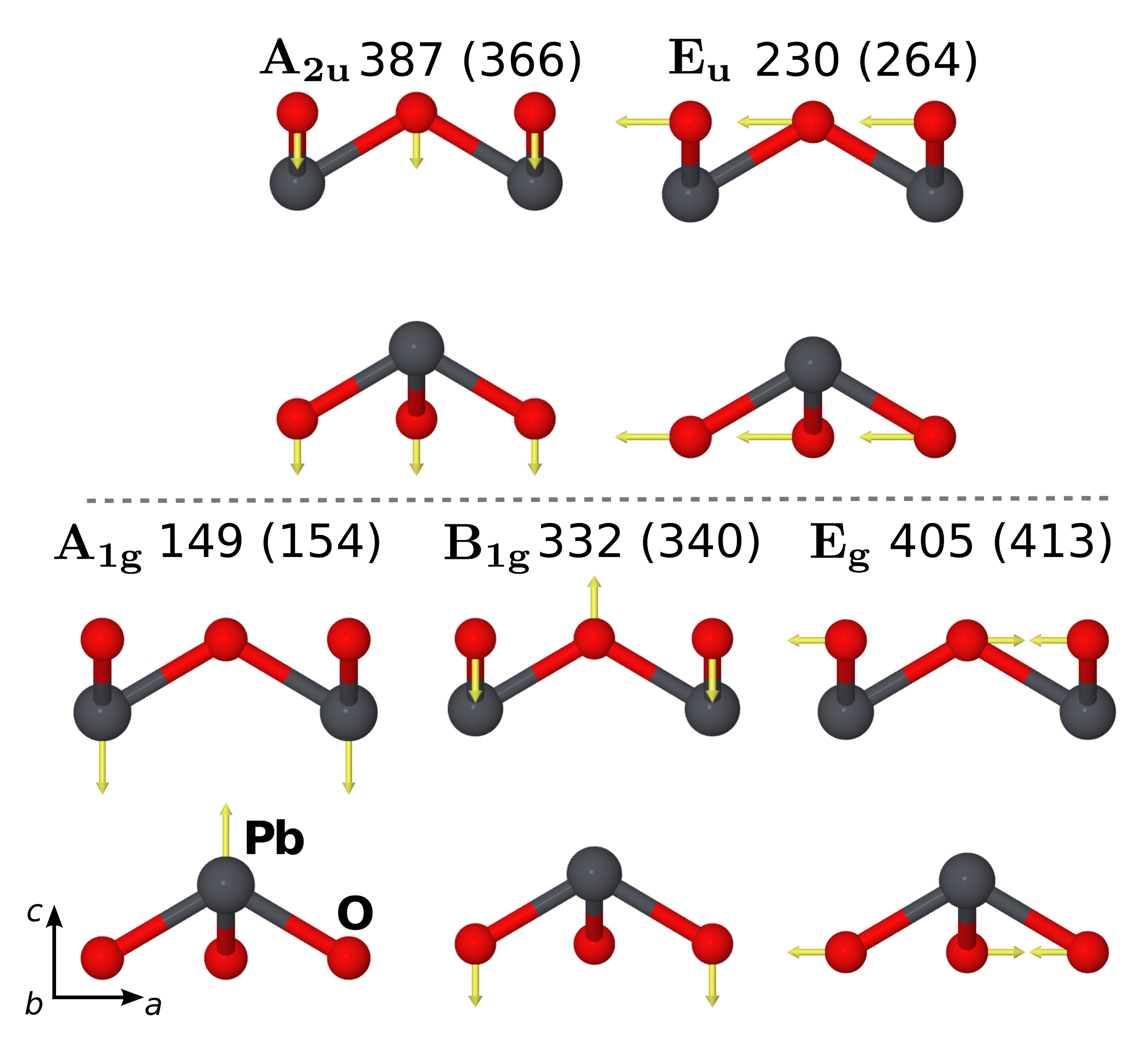}
    \caption{active IR (on the top part) and Raman (on the bottom part) modes for $\alpha$-PbO in cm$\mathrm{^{-1}}$. Only one degeneracy is shown for vibrational modes classified as $E$. PBE-DC2 values in brackets}
     \label{fig:alphamodes}
\end{figure}

Of the nine modes of $\mathrm{\alpha}$-PbO two are IR active ($A_{2u}$ and $E_{u}$); while Raman spectrum consists of four modes ($A_{1g}$, $B_{1g}$ and $2E_{g}$). $E$ modes degenerate showing same atomic displacements, but orthogonal one another. Fig.~\ref{tab:alphafreq} compares the calculated  IR and Raman frequencies with the experimental data and their graphical representation is shown in Fig.~\ref{fig:alphamodes}. 

\normalsize
\begin{table}[!htp]
\centering 
\caption{IR, Raman (R) frequencies, in cm$\mathrm{^{-1}}$, of $\mathrm{\alpha}$-PbO. Simulated intensities are only available for IR modes (in km mol$\mathrm{^{-1}}$). Irep. for irreducible representation, A for activity, Int. for intensity, E for experimental. Experimental IR and Raman frequencies were measured by Adams \textit{et al.}\cite{Adams1976}, while theoretical data is only available for IR.\cite{Waghmare2005} Whenever the experimental data is available $\Delta\nu$ is calculated from this value.} 
 \label{tab:alphafreq}
\begin{tabular*}{1\textwidth}{@{\extracolsep{\fill}}c c c c c c c c c c c} 
\hline
\hline
Irep. & A & \multicolumn{5}{c}{$\mathrm{\nu}$}   & \multicolumn{2}{c}{$\mathrm{\Delta\nu}$}   & \multicolumn{2}{c}{IR Int.}\\
\hline

	- & -     &    PBE & PBE-DC2                                      &  LDA\textsuperscript{\emph{a}}   &  E IR\textsuperscript{\emph{b}}    &   E R\textsuperscript{\emph{b}}            & PBE & PBE-DC2                                                 & PBE & PBE-DC2   \\
\hline
	$E_{g}$ & R & 405 & 413  &  - & - & 321 & 84  & 92  & - & -  \\
	$A_{2u}$ & IR  & 387 & 366  & 399 & 470  & - &  -83  &  -104 & 655 & 810\\
	$B_{1g}$ & R & 332 & 340  &-  & - & 338  & -6  & 2  & - & - \\
	$E_u$ &  IR &  230 & 264  & 275 & 243 & - &  -13 & 21 &  2269 & 2789 \\
	$A_{1g}$ & R & 149 &  154 & -& - & 146  & 3 & 8  & - & -  \\
	$E_{g}$ & R & 79 & 101  &- & -& 81  & 2 & 22  & - & -  \\
\hline
\hline
 \end{tabular*}
 \textsuperscript{\emph{a}}Ref.\cite{Waghmare2005} PP-PW;  \textsuperscript{\emph{b}}Ref.\cite{Adams1976} single crystal specimen at 295 K.
\end{table}

Raman modes $A_{1g}$ and $B_{1g}$ involve only the motion of lead and oxygen atoms parallel to the \textit{c} axis, while the $A_{2u}$ mode is an anti-phase motion of  Pb and O atoms. $E_{u}$ and $2E_{g}$ show atomic displacements in the \textit{ab} planes. Possible overlapping between IR and Raman bands is well documented in the previous literature.\cite{Donaldson1974,Adams1976} For example in the $\alpha$-PbO IR spectrum, mode $A_{2u}$ falls over the $E_{u}$ one forming a broad band around 290\cite{Donaldson1974} and 278 cm$\mathrm{^{-1}}$.\cite{Adams1976} Reflectance IR spectroscopy\cite{Adams1976} successfully resolved the single bands in two distinct peaks:\footnote{Here, only transversal modes are discussed.} \textit{i.e.} 470 cm$\mathrm{^{-1}}$ ($A_{2u}$) and 243 cm$\mathrm{^{-1}}$ ($E_{u}$, see ~\ref{tab:alphafreq}).  A rather large discrepancy is seen for the PBE $A_{2u}$ mode (-83 cm$\mathrm{^{-1}}$), which gets even worse with PBE-DC2 (-104 cm$\mathrm{^{-1}}$). For the $E_u$ mode a much better agreement is seen with some influence of dispersion. In general, the inclusion of disperive interactions via PBE-DC2 does not introduces substantial changes to the IR and Raman modes. As noticed by Ugliengo \textit{et. al.},\cite{Ugliengo2009} there is no direct dispersion correction to the vibrational frequencies as they only change due to a different optimum geometry. Theoretical LDA IR frequencies (see Table~\ref{tab:alphafreq}),\cite{Waghmare2005} agree with our PBE and PBE-DC2 data. Degeneracy occurring for $E_u$ modes make them more intense than the $A_{2u}$ peak as demonstrated by PBE and PBE-DC2 IR intensities. PBE and PBE-DC2 Raman frequencies are in much better agreement with experimental data than the IR ones. This excludes the $E_{g}$ mode,  which suffers a large ipso-chromic shift inverting the experimental order $B_{1g} > E_{g}$. However, the experimental intensity of this mode is very weak.\cite{Adams1976}

\subsubsection{$\mathrm{\beta}$-PbO} 
\begin{figure}[!ht]
	\centering 
	\includegraphics[scale=1.0]{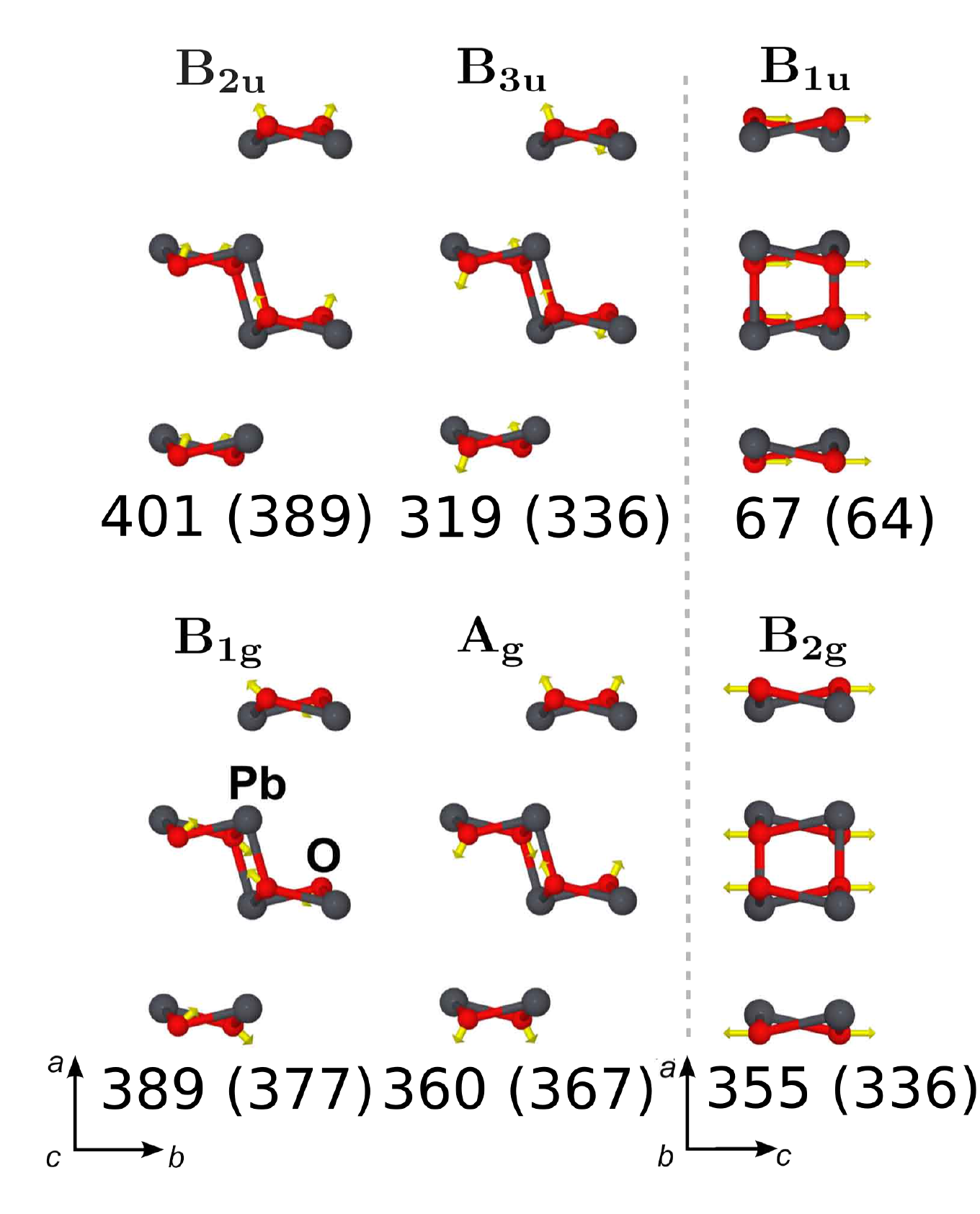}
    \caption{Selection of active IR (on the top part) and Raman (on the bottom part) modes for $\mathrm{\beta}$-PbO in cm$\mathrm{^{-1}}$. Dash line separates modes with different cell-orientation. PBE-DC2 values in brackets. }
     \label{fig:betamodes}
\end{figure}

For $\mathrm{\beta}$-PbO  IR active modes are all those \emph{ungerade} (anti-phase deformation) such as $3B_{2u}$, $3B_{3u}$ and $B_{1u}$, whereas the Raman activities are all \emph{gerade} (in phase deformation) $4B_{1g}$, $2B_{2g}$, $2B_{3g}$ and $4A_{g}$. $2A_{u}$ modes are neither IR nor Raman active, hence they will be not discussed. In Figure~\ref{fig:betamodes} are only shown the graphical atomic displacements of certain modes, which fall at ``high-frequencies'' (500-200  cm$\mathrm{^{-1}}$, exception is $B_{1u}$) \textit{i.e.} $B_{2u}$, $B_{3u}$ for IR, $B_{1g}$, $A_{g}$, $B_{2g}$ and $A_{u}$ for Raman modes.

An electric vector along \textit{b} stimulates the $B_{2u}$ modes, whereas a vector along \textit{c} the $B_{1u}$; both vectors laid normal to the planes formed by the sandwiched arrangement of $\mathrm{\beta}$-PbO (see Fig.~\ref{fig:betamodes} and ~\ref{fig:structures}b). Table~\ref{tab:betafreq} compares the present results with previous experimental works the IR (along with is simulated intensities), Raman frequencies.

\normalsize

\begin{table} [!htp]
\centering 
\caption{IR, Raman (R) frequencies, in cm$\mathrm{^{-1}}$, of $\mathrm{\alpha}$-PbO. Simulated intensities are only available for IR modes and expressed in km mol$\mathrm{^{-1}}$. Irep. for irreducible representation, A for activity, \emph{i} for inactive, Irep. for irreducible representation, Int. for intensity, E is for experimental. Experimental IR and Raman frequencies values were measured by Adams \textit{et al.}.\cite{Adams1976}, while theoretical values are only available for IR.\cite{Waghmare2005} Whenever the experimental data is available $\Delta\nu$ is calculated from this value. } 
 \label{tab:betafreq}
\begin{center}
\begin{tabular*}{1\textwidth}{@{\extracolsep{\fill}}c c c c c c c c c c c  }
\hline \hline
Irep. & A & \multicolumn{5}{c}{$\mathrm{\nu}$} &   \multicolumn{2}{c}{$\mathrm{\Delta\nu}$} & \multicolumn{2}{c}{IR Int.}   \\
\hline
-      & - & PBE & PBE-DC2 &  LDA\textsuperscript{\emph{a}}  &  E IR\textsuperscript{\emph{b}} & E R\textsuperscript{\emph{b}}  &  PBE & PBE-DC2 & PBE & PBE-DC2 \\
\hline
	$B_{2u}$  & IR & 401 & 389 & 418&  356 &   - &  45 & 33 & 1391 & 1913 \\
	$B_{1g}$ & R  & 389 & 377 &  - & - &  385 & 4 & 8& -- & -  \\
	$A_g$ & R & 360 & 367     &  - & - & 349  & 11 & 18 &-&-  \\
	$B_{2g}$ & R & 355 &  336 & - & - & -& - & -& - & -\\
	$B_{2u}$ &  IR & 350 & 337 & 281/360 & 290 & -&  60  & 45 &  65 & 168\\
	$B_{3u}$ & IR & 319 & 336 & 281/360&  424/500 & - &  -181& -165 & 816 & 795    \\
	$B_{1g}$ & R & 310 & 313 &  - &  250 & - & 60 & 63 & - & -\\
	$A_{u}$ & \emph{i} & 293 & 272 &  - & - & - & - & - & - & -\\
	$A_{g}$ & R & 279 & 270  &  - & 289& -& -10 & -19  & - &-  \\
	$B_{3u}$ & IR & 266 & 270 &   281/360 & 424 & - & -158 & -154 & 1219 & 1538  \\
	$B_{3g}$ & R & 222 & 235 &  -  &  171 & -   & 51 &  64 & -     &  -  \\
	$B_{2u}$ & IR & 122 & 136 & - & 146 & -   & -24 & -10   & 0 & 0\\
	$A_g$     &  R &   90 & 100 &  - &   87 & -   & 3    & 13 & - & - \\
	$B_{2g}$ & R &  86 &111&-  &  72 & - & 14 & 39 & - & -\\
	$B_{1g}$ &R &  68 & 74   & - & 52  & - & 16 & 22 & - & - \\
	$B_{1u}$ &  IR &  67 & 64 &  78 & - & -& -11 & -14& 2947 & 3321 \\
	$A_u$ & \emph{i} & 65 & 83 &  - & - &  - & - & - & - & - \\
	$B_{3u}$ & IR & 59 & 64  & - & - & - & - & -& 14 & 21 \\
	$B_{3g}$ & R & 56 & 76 & - & - &  - & -  & - &-& -  \\
\hline
	 
	\hline \hline
 \end{tabular*}
\end{center}
\textsuperscript{\emph{a}}Ref.\cite{Waghmare2005} PP-PW;  \textsuperscript{\emph{b}}Ref.\cite{Adams1976} single crystal specimen at 295 K.
\end{table} 

Adams \textit{et al.},\cite{Adams1976} assigned the main bands to their respective vibrational modes. Simulated frequencies below 68 cm$\mathrm{^{-1}}$ are reported in Table~\ref{tab:betafreq}, however they were not revealed experimentally. PBE and PBE-DC2 IR frequencies are similar to both experimental\cite{Adams1976} and previous theoretical ones.\cite{Waghmare2005} This is particularly true for the $B_{2u}$ mode that suffers a small ipso-chromic shift from the experimental value. Waghmare \textit{et al.} found the same trend using a  PP-PW approach.\cite{Waghmare2005} Puzzling is the  comparison of the $B_{3u}$ modes, which seems underestimated by PBE by 181 cm$\mathrm{^{-1}}$. This mode involves the motion of both Pb and O right across the layered structure (see Fig.~\ref{fig:betamodes}). The PBE-DC2 slightly improve this mode. The correct assignment of the $B_{3u}$ modes is also not very clear  from the experimental point of view, since other modes such $A_{g}$ and $B_{3g}$ would overlap and mix with this modes. Adams \textit{et al.} claimed that these bands could be assigned to overtones.\cite{Adams1976} Waghmare and co-workers addressed this issue reporting a possible spectral window 281-360 cm$\mathrm{^{-1}}$,\cite{Waghmare2005} underestimating the experimental values. Raman frequencies agree very well with those assigned experimentally. The $B_{1g}$ mode suffers of a small up-shift. The other modes fell below this threshold, concluding that our simulation describe the Raman spectrum with good accuracy.

\section{Conclusion}
\label{sec:conclusion}
We demonstrated the use of LCAO approach within the DFT framework to address different properties of lead monoxide polymorphs. 
We tested several GGA and hybrid functionals, in order, to predict as good as possible the PbO geometrical properties. Among the adopted functionals PBE is the best option as Becke's exchange based functionals (BLYP and B3LYP) largely overestimated the cell parameters. The correct geometry is, however, only reproduced when the dispersion interaction is included. In that respect, a new strategy to re-parametrize Grimme's coefficients for Pb and O in PbO is presented, which can be extended also to other semi-ionic solids. This is based on the use of \textit{ab initio} polarizabilities and ionization potentials, which account for the crystalline environments experienced by the ions. The introduction of dispersive interactions was found essential to reproduce the experimental cell parameters for both PbO polymorphs and shown to be the major component of the inter-layer interaction. In accordance with previous computational works our data justify different anisotropy of the Pb lone-pair within the two lead monoxide polymorphs, and this is further confirmed by the quadrupolar coupling constants. Elastic constants clearly show how the $\mathrm{\alpha}$-phase is affected by a larger anisotropy than $\mathrm{\beta}$ one, which eventually reflects the lone-pair orientation within the two PbO-phases. The PBE PbO phonons, at the $\Gamma$-point, for both phases are only in moderate agreement with the experiment and inclusion of dispersion at PBE-DC2 slightly worsen the agreement for the alpha-phase while it remains almost the same for the beta one. As anharmonicity should not play a key role for these systems we suspect that the Grimme's approach to account for  dispersion does not improve frequencies as it has has only an indirect effect through the change in the equilibrium geometry. Further study is needed to clarify this matter.
\acknowledgments
This research was supported by a UKC Scholarship from University of Kent. The authors would like to acknowledge the use of the UK National Grid Service in carrying out this work. PC and PU are very thankful to Dr. Marta Corno and Dr. Bartolomeo Civalleri of the Department of Chemistry, University of Torino. We are finally very thankful to both referees for their comments.

\bibliography{biblio}

\end{document}